\documentclass[twocolumn]{aastex631}
\usepackage{amsmath,amssymb,amsfonts}
\usepackage{algorithmic}
\usepackage{graphicx}
\usepackage{textcomp}
\usepackage{wrapfig}
\usepackage{tabularx}

\usepackage{upgreek}
\usepackage{booktabs}

\usepackage{newtxmath}
\usepackage{multirow}
\usepackage{booktabs}
\DeclareMathAlphabet{\mathfrak}{U}{euf}{m}{n}
\SetMathAlphabet{\mathfrak}{bold}{U}{euf}{b}{n}
\usepackage{amsmath}
\usepackage{graphicx}
\usepackage{amssymb}

\begin{document}

\title{Light-tight skipper-CCDs for X-ray detection in space}
\reportnum{FERMILAB-PUB-25-0963-PPD}

\correspondingauthor{Ana Martina Botti}
\email{botti.anamartina@gmail.com}

\author[0000-0001-6396-2467]{Ana M. Botti}
\affiliation{Fermi National Accelerator Laboratory, PO Box 500, Batavia IL, 60510, USA}

\author{Yikai Wu}
\affiliation{C.N. Yang Institute for Theoretical Physics, Stony Brook University, Stony Brook, NY 11794, USA}
\affiliation{Department of Physics and Astronomy, Stony Brook University, Stony Brook, NY 11794, USA}

\author{Brenda Cervantes}
\affiliation{Fermi National Accelerator Laboratory, PO Box 500, Batavia IL, 60510, USA}

\author{Claudio Chavez}
\affiliation{Fermi National Accelerator Laboratory, PO Box 500, Batavia IL, 60510, USA}

\author{Juan Estrada}
\affiliation{Fermi National Accelerator Laboratory, PO Box 500, Batavia IL, 60510, USA}

\author{Stephen E. Holland}
\affiliation{Lawrence Berkeley National Laboratory, One Cyclotron Road, Berkeley, California 94720, USA}

\author{Nathan Saffold}
\affiliation{Fermi National Accelerator Laboratory, PO Box 500, Batavia IL, 60510, USA}

\author{Javier Tiffenberg}
\affiliation{Fermi National Accelerator Laboratory, PO Box 500, Batavia IL, 60510, USA}
\affiliation{Universidad de Buenos Aires, Facultad de Ciencias Exactas y Naturales, Departamento de Física, Buenos Aires, Argentina}

\author{Sho Uemura}
\affiliation{Fermi National Accelerator Laboratory, PO Box 500, Batavia IL, 60510, USA}



\begin{abstract}

Skipper Charge-Coupled Devices (skipper-CCDs) are pixelated silicon detectors with deep sub-electron resolution. Their radiation hardness and capability to reconstruct energy deposits with unprecedented precision make them a promising technology for space-based X-ray astronomy. In this scenario, optical and near-infrared photons may saturate the sensor, distorting the reconstructed signal. We present a light-tight shield for skipper-CCDs to suppress optical backgrounds while preserving X-ray detection efficiency. We deposited thin aluminum layers on the CCD surface using an e-beam evaporator and evaluated their blinding performance across wavelengths from 650 to 1000 nm using a monochromator, as well as the X-ray transmission using an $^{55}$Fe source. We find that 50 and 100\,nm layers provide $>$99.6\% light suppression, with no efficiency loss for 5.9 and 6.4\,keV X-rays. In addition, we used Geant4 simulations to extend these results to a broader energy range and quantify the efficiency loss for different aluminum thicknesses. Results show that thin aluminum coatings are an effective, low-cost solution for optical suppression in skipper-CCDs intended for X-ray detection and space instrumentation.

\end{abstract}



\section{Light-tight skipper-CCDs} \label{sec:intro}

Skipper Charge-Coupled Devices (CCDs) are silicon pixelated detectors with ultra-low noise. Unlike traditional CCDs, they enable multiple non-disruptive readouts of the same charge packet to achieve deep sub-electron resolution. This  allows determining the charge produced after an interaction in the silicon bulk with unprecedented sensitivity~\cite{janesick_1990,Tiffenberg:2017aac}. Skipper-CCDs are uniquely suited for experiments that require high precision and sensitivity to low-energy signals, such as in neutrino detection~\cite{Atucha} and single-photon imaging~\cite{GhostImaging,QuantumImaging}.

Skipper-CCDs are also suitable for light-dark matter searches. In this scenario, dark matter particles are hypothesized to interact with electrons in silicon, producing only a few electron-hole pairs. To achieve the required sensitivity, we build arrays of skipper-CCDs that add up to a silicon mass ranging from a few grams to a few kilograms~\cite{senseicollaboration2023sensei,2tcc-bqck,aguilar2022oscura}. We deploy them in underground facilities with multiple layers of shielding to attenuate environmental radiation. We also cool them to reduce the dark current. Recently, we demonstrated that infrared light from the blackbody radiation of warm materials in the vessel can leak into the sensors and significantly contribute to the single-electron background, one of the dominant backgrounds in these searches~\cite{sensei1epaper}. Improving the shield near the sensors reduced this background by one order of magnitude, from $\sim10^{-4}$ to $\sim10^{-5}$ electrons/pix/day, but a dedicated shield is still needed to mitigate residual light leaks. The next generation of experiments, such as OSCURA~\cite{OSCURA_Loi}, aims to achieve an additional order-of-magnitude reduction in the single-electron background with a 10\,kg target mass. In this case, detectors will be cooled with nitrogen, and scintillating light will introduce an additional background. In this sense, a dedicated light shield would block both the vessel's infrared emission and the nitrogen's scintillation light.  

In space-based applications, CCDs face a different set of challenges regarding background noise. In this scenario, detectors are exposed to optical photons that may interfere with X-ray signals by saturating the sensor and degrading the energy resolution~\cite{Chandra}. Space-based searches of dark matter may include the detection of keV X-rays originating from dark matter annihilation or decay in the galactic center, as well as single- and few-electron signals from direct sub-GeV dark matter interactions~\cite{ALPINE20254793}. In this scenario, the observing strategy targets regions away from the Sun during umbral phases where the baseline optical background is expected to be zodiacal light~\cite{Leinert}. For a zodiacal light intensity in the visible band of $ 10^{12}\sim10^{13}\ \mathrm{photons}\ \mathrm{m}^{-2}\mathrm{s}^{-1}\mathrm{sr}^{-1}$ and a field of view of $40^\circ$, the expected flux of the optical photon background would be $10^{11}\sim10^{12}\ \mathrm{photons}\ \mathrm{m}^{-2}\mathrm{s}$. Given a pixel pitch of 15\,$\upmu m$, and a frame readout time of 60\,s, the optical diffuse background corresponds to $10^{3}\sim10^{4}$ photons per pixel per frame. To ensure that the detector operates in a regime where the X-ray energy resolution reaches the Fano limit, we required a suppression of the diffuse optical background in the order of $10^{3}\sim10^{4}$, which equals a contribution of $<$1 photon per pixel per frame.

In this work, we present a solution to overcome these challenges: a light shield consisting of a thin aluminum layer deposited directly on the CCD surface, which blocks optical photons while preserving X-ray transmission. This layer can be designed to cover the entire CCD surface for X-ray and dark matter detection, or to cover only a portion to build frame-transfer CCDs to improve the signal-to-noise ratio in space-based astronomy~\cite{tindall, XRISM1, XRISM2, rexis} and imaging applications~\cite{Sierks2011}. In Section~\ref{sec:fab} we describe the fabrication of the aluminum shield, and the testing set-up to assess its blinding performance and X-ray transmission in a front-illuminated sensor. In Section~\ref{sec:calib}, we present the calibration and analysis methods, and in Section~\ref{sec:results}, we present the experimental results. In Section~\ref{sec:sims}, we report on simulations used to determine the X-ray detection efficiency over a broader energy range and in different configurations for a front- and back-illuminated sensor, while in Section~\ref{sec:summary} we conclude with a summary and prospects for future work.

\section{Fabrication and testing set-up} \label{sec:fab}

The first aluminization of a skipper-CCDs was performed at the Center for Nanomaterials at Argonne National Laboratory. We utilized die from a prototype batch of dark-matter sensors~\cite{oscurasensors}; each die consists of a 1.35\,Mpix (1278 $\times$ 1058 pixels) skipper-CCD with four read-out amplifiers located at the corners, and a pixel pitch of 15\,$\upmu$m.

\begin{figure}[t]
    \centering
    \includegraphics[width=1.0\linewidth]{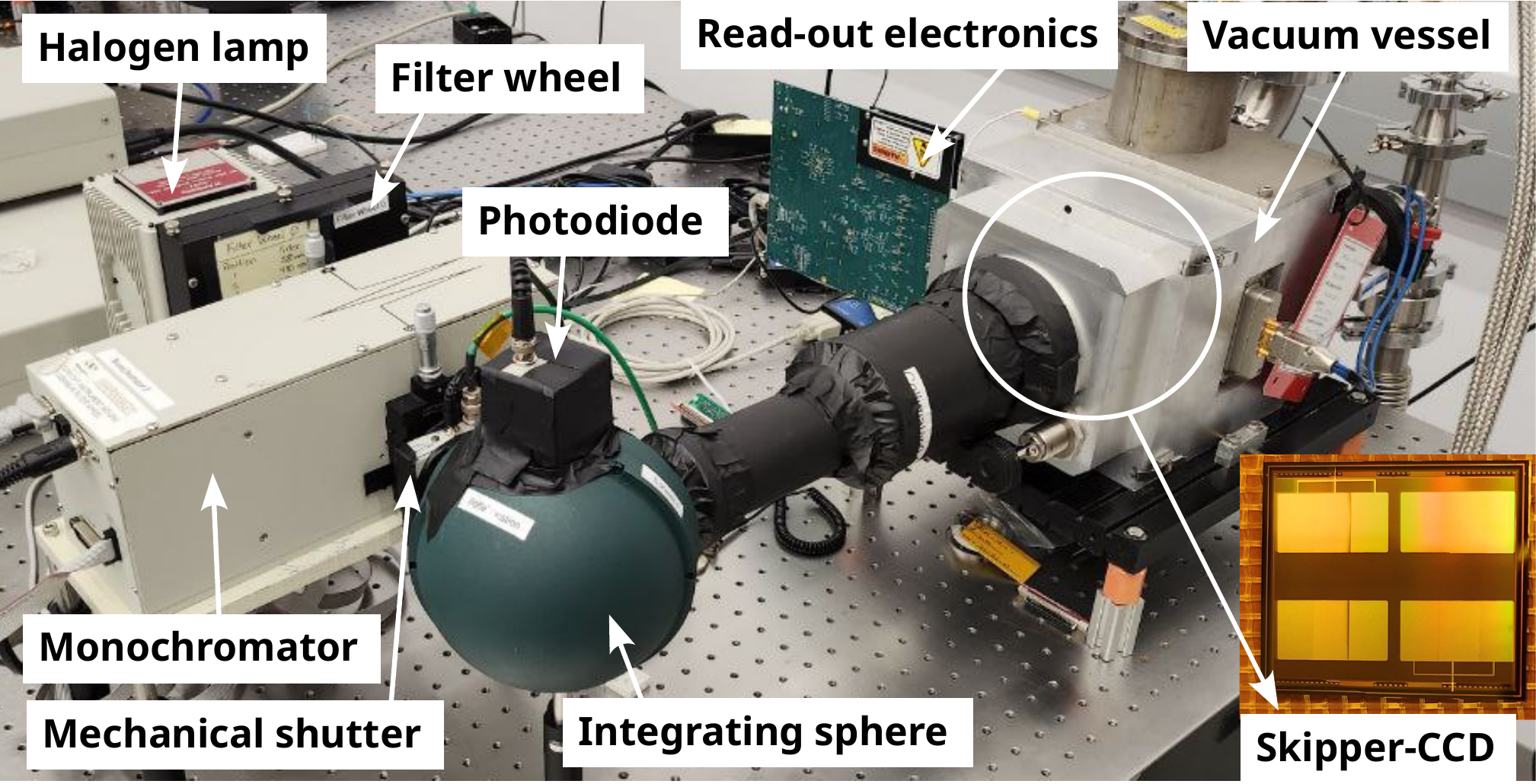}
    \caption{Experimental configuration for testing the skipper-CCD with the aluminum shield. With this setup, we first determined the blinding factor for different wavelengths using a monochromator, and then estimated the X-ray transfer efficiency using a $^{55}$Fe source inside the vacuum vessel.}
    \label{fig:setup}
\end{figure}

To fabricate the aluminum shield, we utilized a liftoff process. We first spin SPR955 photoresist on the frontside of the CCD and use a Heidelberg MLA 150 Maskless Lithography tool to pattern the shield design, which consists of rectangles atop each quadrant, leaving part of the active area uncovered. We developed the photoresist using Microposit$^\mathrm{TM}$ MF$^\mathrm{TM}$-CD-26 developer and deposited aluminum layers of 20\,nm, 50\,nm, or 100\,nm thicknesses using a Temescal FC2000 E-Beam Evaporator. We cleaned the die using Microposit$^\mathrm{TM}$ Remover 1165 and an E3511 ESI Plasma Asher to remove any residue. As a final step, verified the aluminum thickness with a KLA Tencor P7 stylus profiler on a mock die. We show a picture of the CCD with the aluminum shield in the bottom-right corner in Figure~\ref{fig:setup}. The CCD front surface profile comprises an aluminum shield and a dead layer consisting of SiO$_2$, Si$_3$N$_4$, and polysilicon, above the silicon bulk. We summarize the main features of the CCD used in this work in table~\ref{table:CCDspecs}.

\begin{table}
\centering
\begin{tabular}{l l l}
\toprule
Feature & \multicolumn{2}{l}{Value} \\
\midrule
Pixels & \multicolumn{2}{l}{1.35\,Mpix (1278 $\times$ 1058 pixels)} \\
Pixels size & \multicolumn{2}{l}{15\,$\upmu$m} \\
Readout amplifiers & \multicolumn{2}{l}{four skippers} \\
Package type & \multicolumn{2}{l}{Front-illuminated} \\
Light shield & Al & 0, 20, 50, and 100\,nm \\
\multirow{4}{*}{Deadlayer (front)} 
& SiO$_2$ & 1.5\,$\upmu$m \\
& Polysilicon & 250\,nm \\
& Si$_3$N$_4$ & 50\,nm \\
& SiO$_2$ & 50\,nm \\
Bulk & Si & 725\,$\upmu$m \\
 \bottomrule
\end{tabular}
 \caption{Main features of the tested skipper-CCD with aluminum shield. }
 \label{table:CCDspecs}
\end{table}

Regarding packaging, the die is glued to a silicon substrate and wirebonded to a flex cable that connects the CCD pads to the readout electronics. We designed the testing setup to evaluate both the shield's blinding performance and its X-ray detection efficiency. In Figure~\ref{fig:setup}, we show a picture of the testing system, which consists of a monochromator fed with a halogen lamp through a filter wheel to select light wavelength in a range between 300 and 1000\,nm with a resolution $\lesssim$10\,nm. After the monochromator, we installed a shutter and an integrating sphere to ensure isotropic illumination of the sensor, which was located inside a vacuum vessel with a window for light transmission. We pumped the vessel to a pressure below $10^{-4}$\,torr and cooled the sensor to 160\,K with a cryocooler. To control the sensor and extract the video signals, we use a low-acquisition-threshold board~\cite{LTA}. We covered the setup with black and blackout fabric. We do not show the power sources, the Lakeshore PID temperature controller, the vacuum pump, or the crycooler in the image.

The first part of the tests aims to assess the shield's blinding factor. We use different wavelengths to illuminate the sensors and take images with exposures of 0, 5, 10, 30, 60, and 120\,s. In this context, we refer to ``exposure time'' as the time the shutter after the monochromator remains open and the CCD is exposed to the monochromator light; after the exposure, the shutter is closed, and the readout starts. After each illuminated image, we take a control image with the shutter closed to remove potential effects from light leaking into the vessel, dark current, and read-out contributions such as amplifier light and spurious charge~\cite{SEE}. We obtained images with 10 skipper samples per pixel for faster readout and 200 samples for calibration, with a readout noise of 1.22 and 0.25 electrons, respectively.

To evaluate the X-ray transmission through the shield, we installed a $^{55}$Fe radioactive source inside the vessel. We repeated the acquisition sequence with exposures of 0, 120, 300, and 600\,s, but with the monochromator shutter closed. In this case, the CCD was also exposed to X-rays during readout.

\begin{figure}[t]
    \centering
    \includegraphics[width=1.0\linewidth]{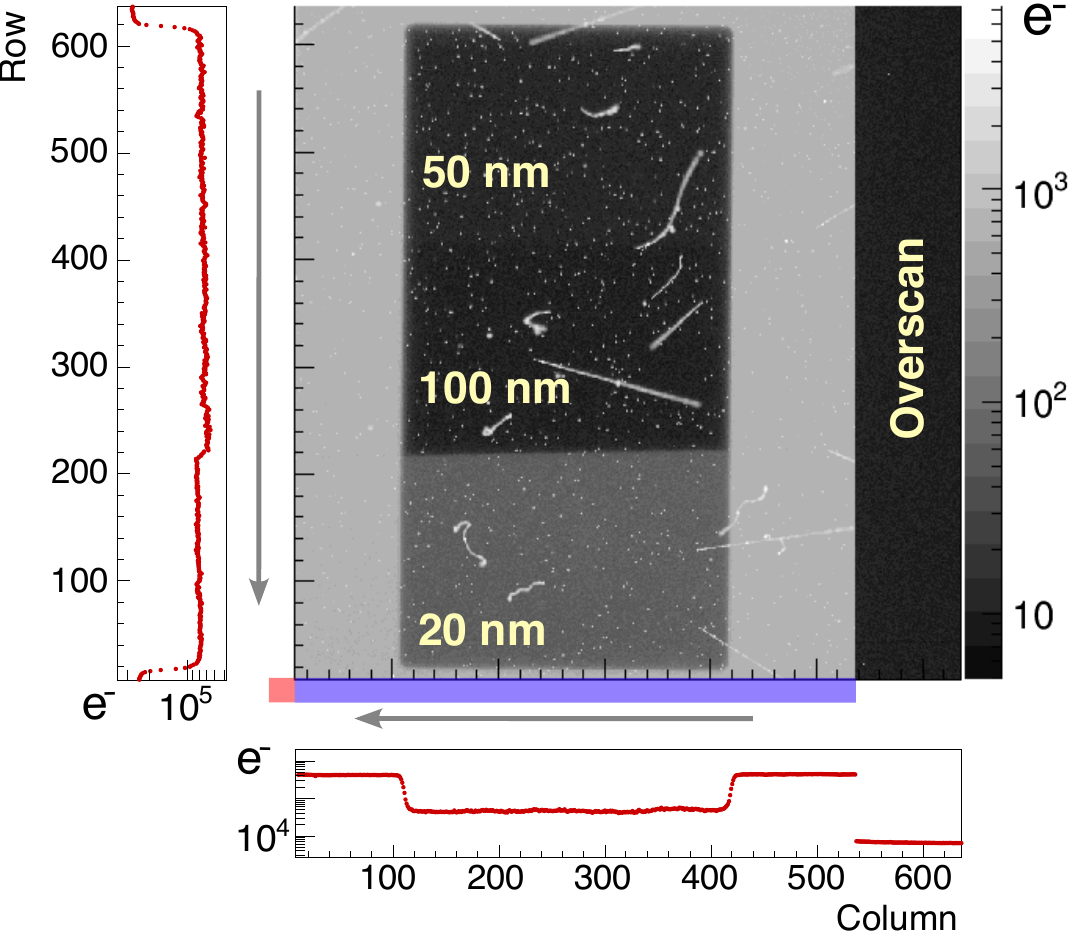}
    \caption{Image obtained with one quadrant of the skipper-CCD. We illuminate the sensor with 950\,nm photons and the $^{55}$Fe radioactive source. The x(y)-axis in the image correspond to the columns (rows), the blue rectangle represents the position of the serial register, and the red square the readout amplifier. Gray arrows indicate the direction of the read-out. We show the projections on the rows and columns after removing the high-energy events from the radioactive source and environmental radiation. Dark regions correspond to the overscan and aluminum shields of 20, 50, and 100\,nm. 
    }
    \label{fig:img}
\end{figure}

In Figure~\ref{fig:img}, we show an example of a one-quadrant image obtained with the radioactive source inside the vessel, and the monochromator set to 950 nm. The x-axis corresponds to the column number, while the y-axis corresponds to the row number. The blue rectangle indicates the position of the readout stage or ``serial register’’, while the red square represents the readout amplifier. Gray arrows indicate the direction of the readout: after a row is transferred to the serial register, the amplifier senses the signal from each pixel. For each row, we obtained about 100 overscan pixels, which we used to subtract the signal baseline. The gray scale indicates the number of charges collected in the readout, and the plots on the left and bottom correspond to projections along the row and column axes after removing high-energy events. We also show x- and y-axis projections computed by adding the charge of all pixels in the same column and row, respectively. 
Regions not covered by the aluminum shield exhibit high charge levels, while regions covered by the 20\,nm shield partially allow light transmission. Pixels under the 50 and 100\,nm shields exhibit signal levels apparently similar to those in the overscan. Point-like high charge clusters correspond to the $^{55}$Fe X-rays, while long traces and curly traces correspond to muons and electrons, respectively. The column projection shows slopes of the transition between the unshielded and shielded regions, a product of the charge diffusion.

\section{Calibration and data analysis} \label{sec:calib}

\begin{figure}[t]
    \centering
    \includegraphics[trim={0 15cm 0 0}, clip, width=1.0\linewidth]{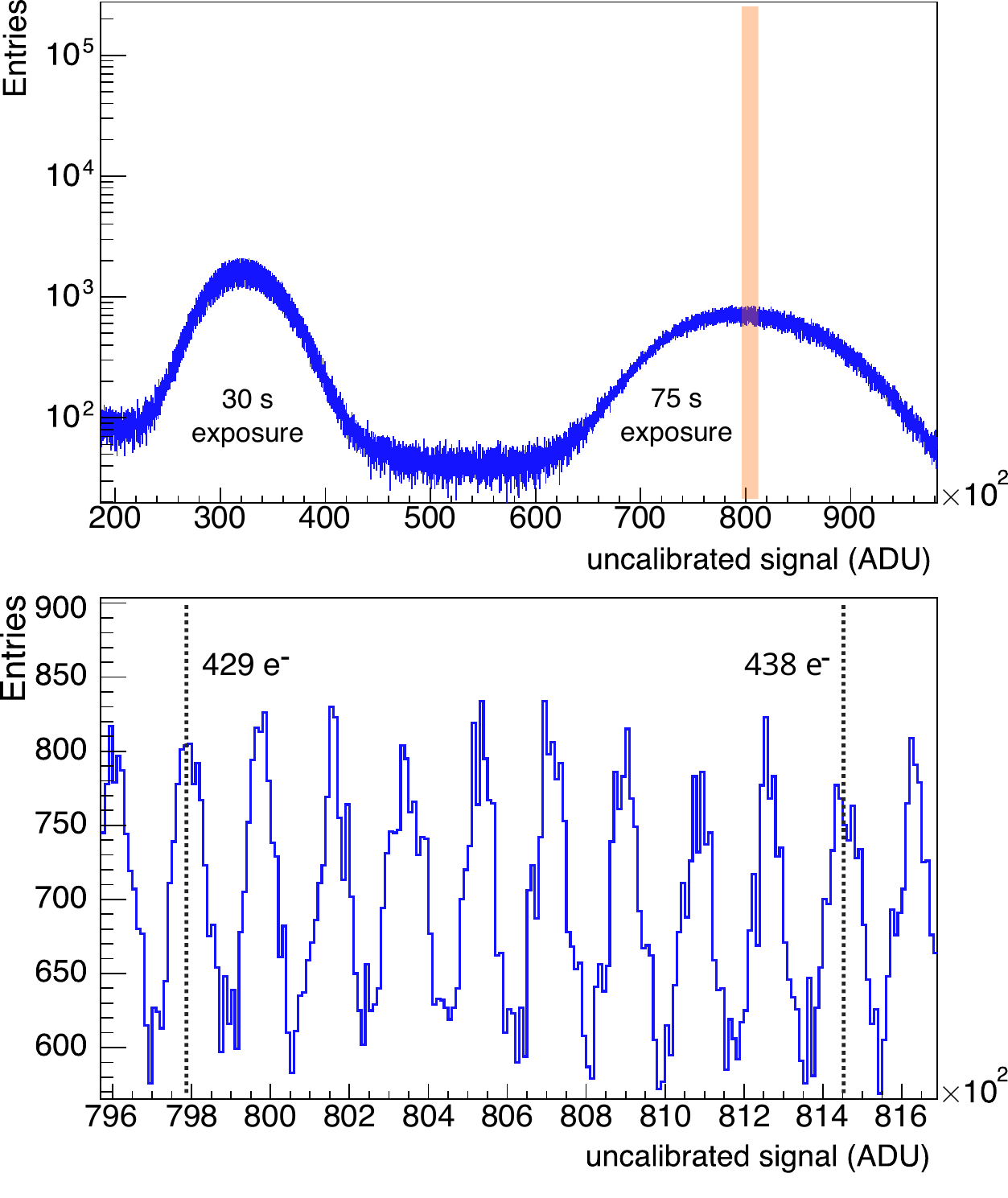}
    \caption{(Top) uncalibrated charge spectrum for images with different exposures to 950\,nm light obtained with one amplifier and 200 samples. (Bottom) Zoom-in of the orange region on top, showing photoelectron peaks between 428 and 439 electrons.}
    \label{fig:calib}
\end{figure}

After each acquisition, we average the 10 or 200 samples per pixel and per image, and subtract the signal baseline calculated from pixels in the overscan region. To calibrate, we used images with 200 samples and computed the histogram of the pixel signal for different exposures to obtain the charge spectrum. In the top panel of Figure~\ref{fig:calib}, we show the pixel charge spectrum, consisting of a histogram in which each entry corresponds to the measured charge of one pixel in one image. We used pixels read by the same amplifier in 18 images with 0\,exposure, 9\,images with 30\,s, and 9 images with 75\,s, illuminated with 950\,nm light; the two bumps in the histogram correspond to the 30 and 75\,s exposures used to produce it. In the bottom panel, we present a zoom-in of the orange region shown in the top panel, where peaks corresponding to different numbers of electrons are visible. We then compute the mean signal level as a function of the peak (or electron) number, from which we determine the number of electrons corresponding to the signal in analog-to-digital converter (ADC) units. It is worth noting that although these images were obtained with the X-ray source installed inside the vessel, the presented spectrum is dominated by charge produced by monochromator light.

To assess the shielding effect, we used images acquired with 10 skipper samples, which were sufficient to reduce the readout noise below the shot noise level. We compared the charge in regions beneath the shield with that in regions without the shield. We confirmed uniform illumination across the quadrant by comparing the signal distribution across different regions. We used control images obtained with the shutter closed to subtract the signal from any possible light leaking into the vessel and light emitted by the vessel walls. We computed the charge in each region as a function of exposure time for different wavelengths. The number of electrons produced per second of exposure was determined by a linear fit. The slope gives the number of charges produced exclusively by the monochromator light, removing those from environmental photons or readout effects. Finally, we calculated the ratio of the signal in the shielded region to the signal in the unshielded region to obtain the blinding factor. High-energy events from environmental radiation, such as muons, were removed from this analysis. 

\begin{figure}[t]
    \centering
    \includegraphics[trim={0 0cm 0 25cm}, clip, width=1.0\linewidth]{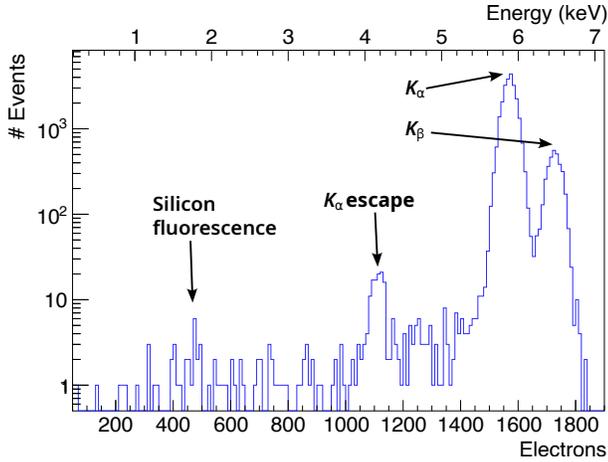}
    \caption{$^{55}$Fe X-ray event spectrum obtained using one quadrant images with 10 samples. The photoabsorption peak for 5.9 (K$_\alpha$) and 6.4\,keV (K$_\beta$) X-ray appears along with the K$_\alpha$ escape and silicon fluorescence peaks.}
    \label{fig:xray}
\end{figure}

We reconstructed the X-ray events using a clustering algorithm that identifies consecutive non-empty pixels and groups them into one event. In this case, we used images with 10 skipper samples per pixel, but with the shutter always closed to avoid the pile-up between the monochromator light and the X-rays. By adding the charge from all pixels in the event, we obtain the event energy, as shown in Figure~\ref{fig:xray}. To convert the energy units from electrons to eV, we used a mean electron-hole pair creation energy of 3.75\,eV~\cite{Rodrigues2020}. The photoabsorption peaks corresponding to the $^{55}$Fe X-ray emissions at 5.9 (K$_\alpha$) and 6.4 (K$_\beta$)\,keV are distinguishable, as well as the K$_\alpha$ escape and silicon fluorescence peaks. 

We identified X-ray events as clusters with charge between 1450 and 1800 electrons, including pile-up events that produce clusters with twice or three times the X-ray energy. In principle, this approach removes events at lower energy produced by X-ray scatterings, such as Compton scattering; however, at this photon energy, interactions are dominated by photoabsorption, and scatterings can be neglected~\cite{fernandezmoroni2020charge}. In addition, we focused on space-based applications in which the search centers on the X-ray absorption peak, such as the detection of spectral lines from dark matter decay or annihilation~\cite{ALPINE20254793}, where the signal is more likely to dominate over the continuum background. 

To estimate the efficiency loss through the aluminum shield, we count X-ray events in pixels with and without the shield in nearby regions to mitigate any spatial non-uniformity. In this case, the CCD remains exposed during the readout, and shielded pixels spend some of the readout time under the unshielded region. We remove the readout contribution by computing the number of events as a function of exposure time and, once again, obtaining the slope from a linear fit. 

Note that fluctuations in the energy reconstruction due to charge-transfer inefficiencies, background pile-up, and the clusterization threshold may degrade the energy resolution, in addition to the readout and Fano noise. Determining the skipper-CCD operation parameters and optimal reconstruction to improve the X-ray energy resolution is beyond the scope of this work. We also note that different acquisition conditions, such as the number of samples, may introduce calibration shifts, which are negligible in this case~\cite{Drlica_2020}. Since this work relies solely on counting events in the X-ray peaks, the energy calibration and its precision do not affect the results.

\section{Blinding and X-ray transmission} \label{sec:results}



To assess the aluminum blinding power, we compared the signal in shielded and unshielded regions. We verified that the signal across different regions without the shield is uniform by comparing their pixel charge distribution. At wavelengths below $\sim$450\,nm, the light is absorbed in the front-illuminated CCD surface structures, which consist of SiO$_2$, Si$_3$N$_4$, and polysilicon layers before the silicon bulk. To detect photons at shorter wavelengths, the CCD needs to be thinned and backilluminated~\cite{Groom1999,Holland:2003}. On the other hand, at higher wavelengths, as the photon energy approaches the silicon bandgap, the CCD becomes transparent.

\begin{figure}[t]
    \centering
     \includegraphics[width=1.0\linewidth]{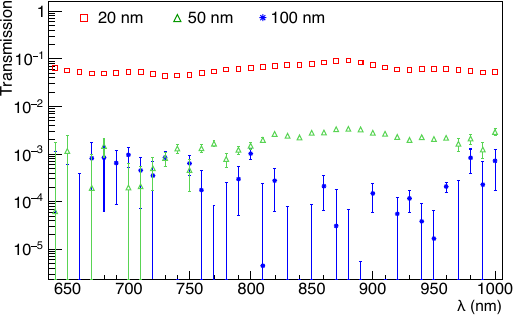}
    \caption{Light transmission factor through the aluminum shield as a function of wavelength for diﬀerent aluminum thicknesses. We calculate the transmission as the ratio of the signal in unshielded and shielded pixels.}
    \label{fig:blinding}
\end{figure}

We present in Figure~\ref{fig:blinding} the light transmission factor as a function of the wavelength for aluminum thicknesses of 20 (red squares), 50 (green triangles), and 100\,nm (blue stars). The transmission is computed as  $\mathrm{\frac{S_{unshielded}}{S_{shielded}}}$, where $\mathrm{S_{shielded}}$ is the signal in the shielded pixels and $\mathrm{S_{unshielded}}$ that in unshielded pixels. 
The aluminum shield clearly attenuates the monochromator light, with stronger attenuation for thicker aluminum layers: the 50 and 100\,nm shields block over 99.6\% and 99.9\%, respectively, of the light across most of the frequency range. In contrast, the 20\,nm shield is not thick enough to block all light, allowing more than 5 to 10\% of it into the sensitive region of the sensor. 
For wavelengths longer than 950\,nm, we used only images with 5 and 10\,s exposures, since photons interact closer to the CCD backside and, at high levels of illumination, diffused charges flood into the shielded region.

We estimated the optical transmission through free-standing aluminum films using tabulated constants~\cite{opticTable} and film thicknesses of 20 nm, 50 nm, and 100 nm over the wavelength range 650 to 1000\,nm. The optical models predict a strong suppression of transmission with the film thickness:  approximately (3.5$\sim$14)\% for 20 nm, (0.1$\sim$0.8)\% for 50 nm, and $<$0.007\% for 100 nm. We measured a transmission of (6.367$\pm$0.006)\% for the 20\,nm film, (0.227$\pm$0.002)\% for the 50\,nm film, and (0.006$\pm$0.002)\% for the 100\,nm film; results are reported as the mean over the 650 to 1000\,nm wavelength range. These results show that the effective aluminum thickness in the tested sensor is consistent with tabulated values, although higher-order deviations due to the surface roughness or the presence of additional layers in the gate structure have a non negligible contribution that is difficult to estimate.

\begin{figure}[t]
    \centering
    \includegraphics[width=1.0\linewidth]{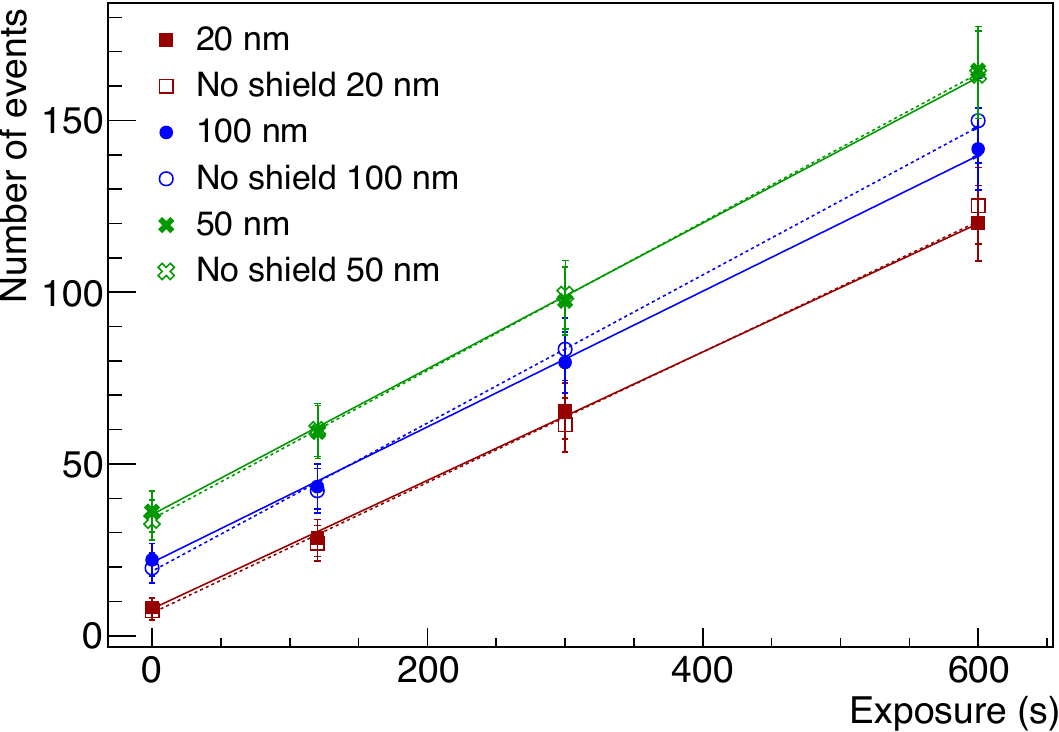}
    \caption{Number of X-ray events per image as a function of exposure time for different regions of the CCD, each region with the same size and number of pixels. The difference in the number of events across regions corresponds to X-rays hitting the sensor during readout. This vertical offset corresponds to the 76-second time to read each region. Lines correspond to the linear fits for the shielded (solid) and unshielded (dashed) regions. We show the fit results in Table~\ref{table:results}.}
    \label{fig:nroXrays}
\end{figure}

We estimated the X-ray detection efficiency loss due to the shield by comparing the number of events per second of exposure between the shielded and unshielded regions. In Figure~\ref{fig:nroXrays} we present the number of events per image as a function of the exposure times for the regions with the shield (full markers) and nearby regions without the shield (open markers); all regions have the same size and number of pixels. The difference in the number of events for the 20 (red squares), 50 (green crosses), and 100\,nm (blue circles) regions corresponds to the events produced during readout: at higher row number, the pixels remain exposed for a longer time.   

In Figure~\ref{fig:nroXrays}, we used linear fits to extract the number of events per image per second of exposure; results are presented in Table~\ref{table:results}. Within the uncertainties, there is no apparent attenuation of the X-ray intensity due to the Aluminum shield. It is worth noting that we did not introduce a correction to the exposure time corresponding to the readout, which increases linearly with the row number. For a readout time of 250\,s, this is about 0.38\,s per row. The vertical offset in the intercept corresponds to the events that hit the sensor during the $\sim$76\,s (200 rows) difference in readout time between regions. When comparing shielded and unshielded regions, we used nearby regions in the same rows to ensure equal readout times.


\begin{table}
\centering
\begin{tabular}{lc}
\toprule
 Shield &  Slope (events / s) \\
 \midrule
 20 nm &  0.19 $\pm$ 0.02 \\
 No shield 20 nm & 0.19 $\pm$ 0.02 \\
 100 nm & 0.20 $\pm$ 0.02  \\
 No shield 100 nm & 0.22 $\pm$ 0.02   \\
 50 nm & 0.22 $\pm$ 0.02\\
 No shield 50 nm & 0.22 $\pm$ 0.02  \\
 \bottomrule
\end{tabular}
 \caption{Slope for linear fits in Figure~\ref{fig:nroXrays}. The number of events per image per second remains unchanged across the CCD, indicating that the aluminum shield does not introduce an apparent efficiency loss for X-rays at 5.9 and 6.4\,keV.}
 \label{table:results}
\end{table}

\section{Simulated detection efficiency vs. X-ray energy} \label{sec:sims}

 We implemented a Geant4~\cite{geant4, Geant42, Geant43} simulation using the Penelope low-energy electromagnetic models with aluminum thicknesses of 20, 50, 100, and 1000\,nm on top of the CCD surface and X-ray energies between 0.1\,keV and 25\,keV. For the X-ray source, we modeled an infinite plane far from the sensor, emitting X-rays with an isotropic angular distribution; the detection efficiency depends on the penetration depth and, in turn, on the incidence angle. We implemented a thick shield layer of 1000\,nm as and extreme case that could serve as a low-energy X-ray filter.

\begin{figure}[t]
    \centering
    \includegraphics[width=1.0\linewidth]{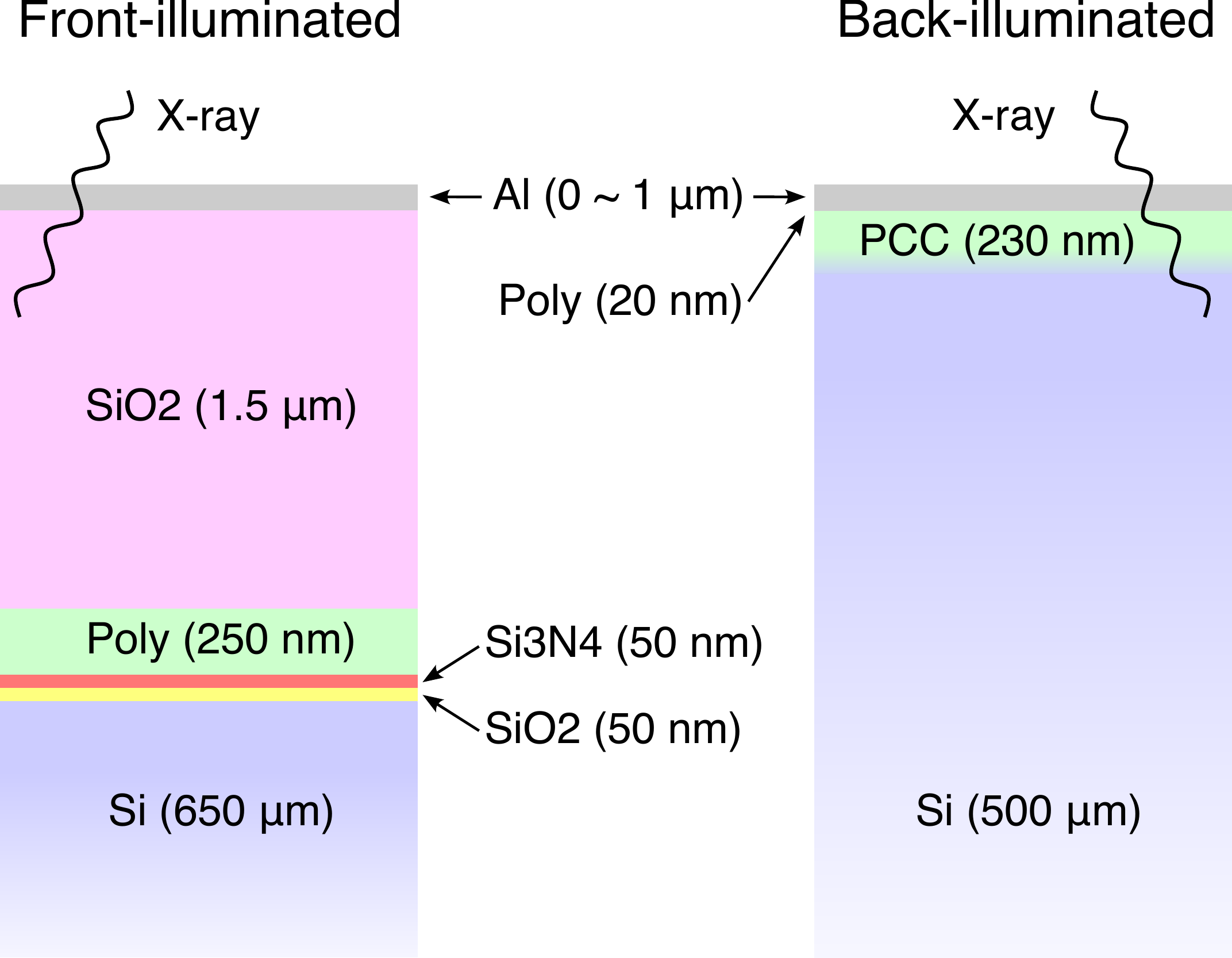}
    \caption{Schematics of the skipper-CCD surface cross sections implemented in the Geant4 simulations. (Left) Frontside. (Right) Backside after thinning and treatment. PCC refers to the partial charge collection layer.}
    \label{fig:schematics}
\end{figure}

\subsection{Front-illuminated} 

We used measurements of the CCD composition profile to simulate X-rays at different energies and estimate the efficiency loss produced by the aluminum shield. We present a schematic of the CCD frontside cross section in the left panel of Figure~\ref{fig:schematics}. In addition to the 650$\sim$720\,$\upmu$m silicon bulk, the CCD front consists of a 1.5\,$\upmu$m SiO$_2$ layer and 0.25\,$\upmu$m of polysilicon corresponding to the clock structures. Between the clocks and the silicon bulk, there is an insulator consisting of 50\,nm Si$_3$N$_4$ and 50\,nm SiO$_2$ layers.

\begin{figure}[t]
    \centering
    \includegraphics[width=1.00\linewidth]{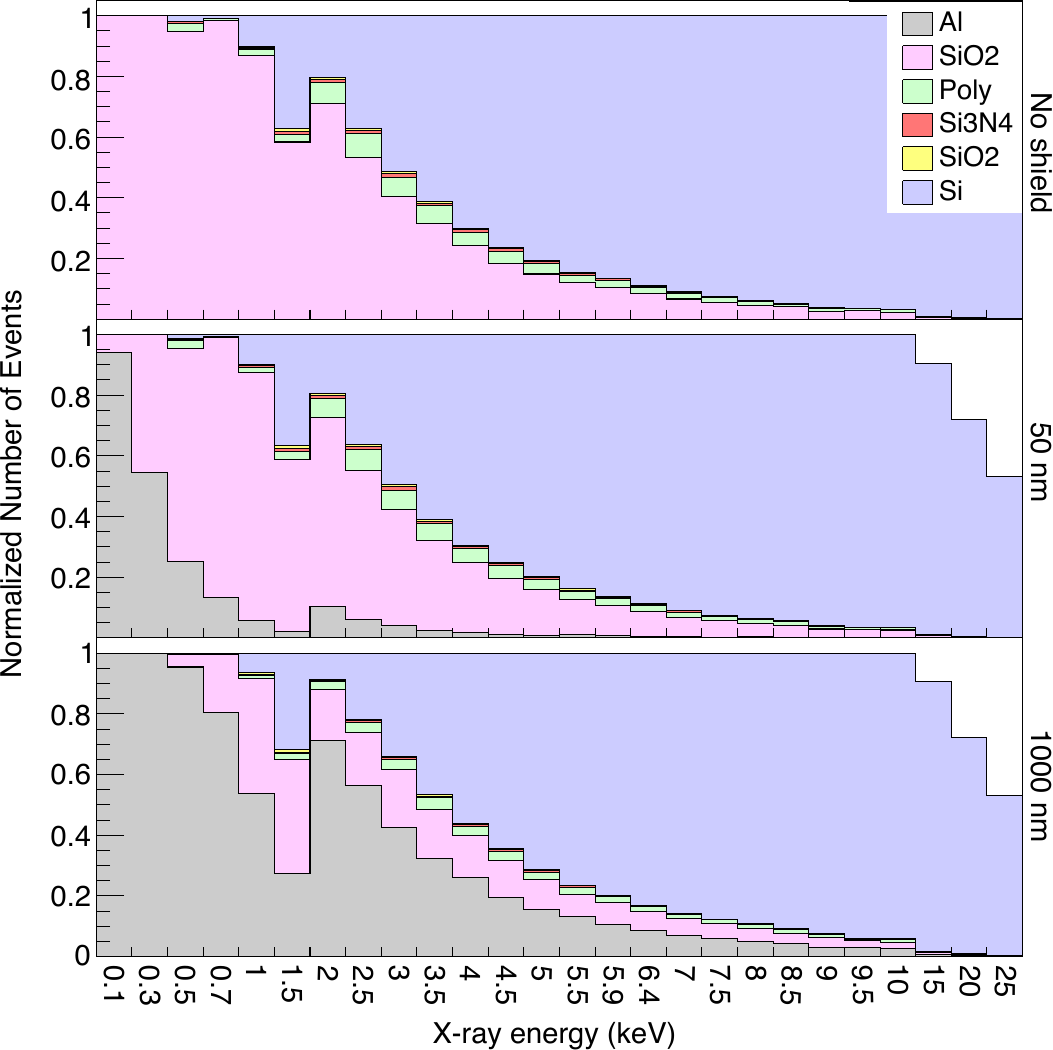}
    \caption{Normalized number of events interacting in the different layers of the CCD front surface for a sensor without a shield (top panel), a 50\,nm (middle), and a 1000\,nm (bottom) shields. We present results for different X-ray energies; in each bin, we construct a stack of events interacting in each region.}
    \label{fig:nroEventsFront}
\end{figure}

In Figure~\ref{fig:nroEventsFront}, we present the normalized number of events interacting in the different layers of the CCD for each X-ray energy. We removed events that impinged on the sensor edges, since in a realistic application, we would have a shield around the CCD. We present results with no aluminum shield (top panel), 50 (middle panel), and 1000\,nm (bottom panel) shield. For each energy, we show a stack of the events interacting in the aluminum (gray), SiO$_2$ (pink), polysilicon (green), Si$_3$N$_4$ (red), second SiO$_2$ (yellow), and silicon (blue) layers. We estimated the number of events in each layer using both the position of the first interaction and the volume in which more than 90\% of the X-ray energy is deposited to corroborate that they are the same. This is expected, since photoabsorption is the dominant process at these energies. Note that the x-axis is not a linear scale.

At lower energies, X-rays are attenuated mainly at the SiO$_2$ layer, and the aluminum shield does not produce a significant difference in the number of events penetrating into the silicon bulk. As the photon energy increases, the attenuation length increases, leading to more interactions in silicon. The jump between 1.5 and 2\,keV corresponds to the increase of the interaction probability when X-ray energies exceed the silicon and aluminum K-shell energies of about 1.84 and 1.56\,keV, respectively. At energies above 9.5\,keV, the attenuation length approaches the sensor thickness, and X-rays begin to traverse the CCD without interacting. The 50\,nm aluminum shield stops less than 5\% of the X-rays with 3.5\,keV energy, while the 1000\,nm shield stops more than 30\% at this same energy.

\begin{figure}[t]
    \centering
    \includegraphics[width=1.0\linewidth]{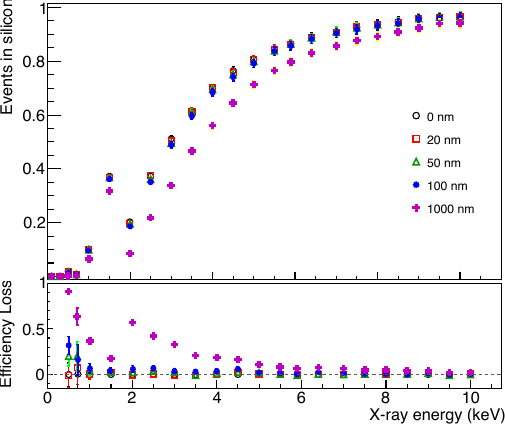}
    \caption{(Top) Ratio between the number of events interacting in the silicon bulk and the total number of events interacting in a front-illuminated CCD as a function of the X-ray energy for different aluminum shield thicknesses. (Bottom) Efficiency loss due to the aluminum layer as a function of the X-ray energy.}
    \label{fig:effFront}
\end{figure}

To better illustrate the efficiency loss due to the aluminum shield, we present in the top panel of Figure~\ref{fig:effFront} the ratio between the events interacting in the silicon bulk and total number of events in the sensor ($\mathrm{\frac{N_{si}}{N_{total}}}$) as a function of the X-ray energy. We show results for the 20 (red squares), 50 (green triangles), 100 (blue full circles), and 1000\,nm (purple crosses) aluminum layers, as well as for no shield (open black circles). In the bottom panel of Figure~\ref{fig:effFront} we report the effiency loss due to the aluminum shield as a function of the X-ray energy, calculated as $\mathrm{1-\frac{N_{si}}{N^{No\ shield}_{si}}}$, where $\mathrm{N_{si}}$ and $\mathrm{N^{No\ shield}_{si}}$ is the number of events interacting in the silicon for the detector with and without shield, respectively. For aluminum layers thinner than 100\,nm, we do not observe a significant efficiency loss due to the shield; the dominant loss is caused by attenuation in the SiO${_2}$ and polysilicon layers. These results are consistent with the laboratory measurements reported in Figure~\ref{fig:nroXrays}. A thicker aluminum shield could be useful in a scenario where, in addition to optical photons, the background is also composed of lower-energy X-rays.

\subsection{Back-illuminated} 

Typically, in astronomical applications, CCDs are thinned and back-illuminated to improve the photodetection efficiency at low wavelengths. In the X-ray regime, the structures on the frontside reduce the X-ray detection efficiency to 60$\sim$65\% at 3.5\,keV, which motivates using a thinned back-illuminated sensor. The thinning and backside treatment in this case would aim to minimize the thickness of the dead layer on the CCD's backside. 

To increase the detection efficiency of low-energy X-rays, we could implement the same backside processing used in fully depleted astronomical CCDs~\cite{Holland:2003}, but without antireflective coatings. Related fabrication efforts have already demonstrated the production of 580\,$\upmu$m-thick CCDs with a thin backside in situ doped polysilicon and no antireflective coatings~\cite{grailCCD}. After the wafer thinning and polishing of the back surface of the CCD, a layer of in-situ doped (phosphorus) polysilicon is deposited to create the backside ohmic contact. However, as a consequence of this process, a partial charge collection layer is also formed, which consists of a transition region between the polysilicon and the silicon bulk with a gradient of phosphorus concentration~\cite{fernandezmoroni2020charge}. Charges produced in this region will be partially recombined, and only a fraction will be collected. This will distort the reconstruction of X-ray event energy, potentially leading to a loss in efficiency. 

We implemented a Geant4 simulation to study the performance of back-illuminated Skipper-CCD; in this scenario, we simulated the sensor backside using previous estimates of the polysilicon dead layer (20\,nm) and the partial charge-collection layer ($<$230\,nm). We adopted a conservative approach, assuming that any interaction in these regions is lost. We simulated the same sensor and source geometries, the same X-ray energies, and aluminum thicknesses as in the front-illuminated case. We reduced the silicon thickness to 500\,$\upmu$m to account for the thinning. We present a schematic of the backside in the right panel of Figure~\ref{fig:schematics}. It is worth noting that thinner partial charge-collection layers have been demonstrated in various instruments~\cite{backside1,backside2}, and we present a conservative estimate of the X-ray detection efficiencies based on measurements performed with skipper-CCDs~\cite{pcc, pcc2}.

\begin{figure}[t]
    \centering
    \includegraphics[width=1.0\linewidth]{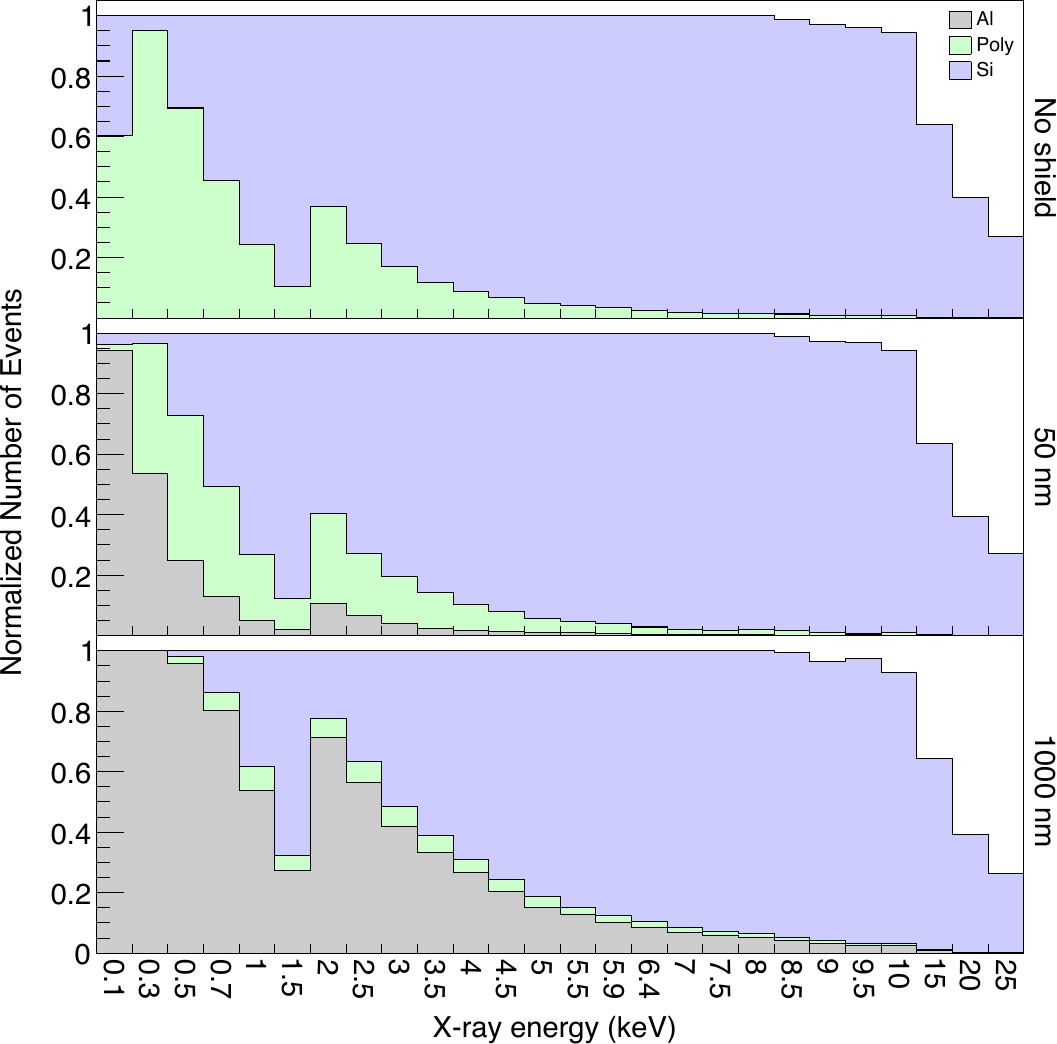}
    \caption{Normalized number of events interacting in the different layers of the CCD back surface for a sensor without a shield (top panel), a 50\,nm (middle), and a 1000\,nm (bottom) shields. We present results for different X-ray energies; in each bin, we construct a stack of events for each region.}
    \label{fig:nroEventsBack}
\end{figure}

In Figure~\ref{fig:nroEventsBack}, we present the normalized number of events interacting in each of the backside layers for different X-ray energies. The top, middle, and bottom panels show results for a CCD without a shield, and with 50\,nm and 1000\,nm shields, respectively. We treat the polysilicon and partial charge-collection layers as a single dead layer. Since the silicon bulk is thinner in this case, the efficiency at higher energies falls faster than in the front-illuminated case. In the unshielded sensor, a new step at 0.1\,keV appears, corresponding to the first silicon L-shell. It is worth noting that standard Geant4 libraries do not accurately simulate interactions at these low energies, and results are only qualitative~\cite{skipper_compton, DAMIC-M:2022xtp}.

\begin{figure}[t]
    \centering
    \includegraphics[width=1.0\linewidth]{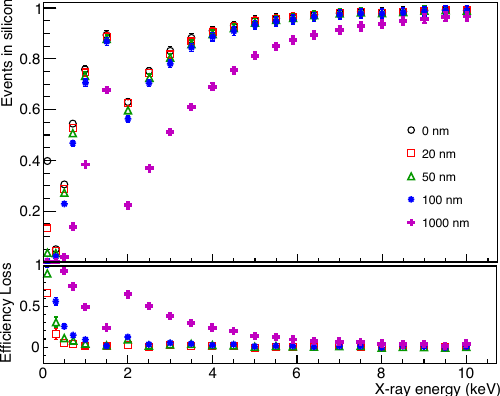}
    \caption{(Top) Ratio between the number of events interacting in the silicon bulk and the total number of events interacting in a back-illuminated CCD as a function of the X-ray energy for different aluminum shield thicknesses. (Bottom) Efficiency loss due to the aluminum layer as a function of the X-ray energy.}
    \label{fig:effBack}
\end{figure}

Similar to the front-illuminated simulation, we present in the top panel of Figure~\ref{fig:effBack} the fraction of events interacting in the silicon bulk as a function of the X-ray energy. The thinner dead layer at the CCD surface allows for a significantly higher detection efficiency. At 3.5\,keV the efficiency is increased to above 85\%. It should be noted that these efficiency estimates do not account for losses due to data selection, pile-up, and reconstruction, which are beyond the scope of this work. 

In the bottom panel of Figure~\ref{fig:effBack}, we present the efficiency loss due to the aluminum shield. For energies above 1\,keV, the efficiency loss due to the aluminum shield is consistent with results from the front-illuminated case, since the X-ray energy is sufficient to penetrate both dead layers. However, at lower energies, the thinner dead layer in the back-illuminated case allows the detection of a fraction of X-rays, some of which can be stopped by the aluminum. This efficiency loss becomes more significant for X-rays with energy below 0.5\,keV. At higher energies, the detection efficiency is conserved for shields thinner than 100\,nm.

\section{Summary} \label{sec:summary}

In this work, we developed and tested thin aluminum light shields deposited directly on the surface of skipper-CCDs to suppress optical backgrounds while preserving X-ray detection efficiency. We fabricated aluminum layers of 20, 50, and 100\,nm using an e-beam evaporator. This approach was suitable for this proof of concept; however, since e-beam evaporation may produce radiation damage, other fabrication methods should be implemented in future iterations. Several alternatives, such as thermal evaporation, are widely available and fabrication should not represent a technical challenge~\cite{rexis}.

Using monochromatic illumination, we evaluated the light transmission for wavelengths between 650 and 1000\,nm and found that the 50 and 100\,nm coatings block more than 99.6\% and 99.9\% of incident photons, respectively, in most of this range. The 20\,nm layer does not provide adequate suppression, transmitting 5$\sim$10\% of the incoming light across a broad spectral region. For wavelengths below 650 nm, photons are stopped by structures on the CCD surface, and the transmission cannot be assessed without a back-illuminated sensor. Nevertheless, the aluminum reflectiveness is quite uniform in the 300$\sim$650\,nm range, and similar results are expected. For wavelengths above $\sim$1000\,nm, the CCD becomes transparent as the photon energy approaches the silicon bandgap. 

The X-ray transmission measurements using a $^{55}$Fe source indicate that the aluminum layers did not introduce an apparent efficiency loss at 5.9 and 6.4\,keV. To validate these results, we implemented a Geant4 simulation of the CCD front structure, consisting of an aluminum shield and SiO$_2$, polysilicon, Si$_3$N$_4$, and silicon layers. We simulated X-rays isotropically impinging the CCD front, showing that the efficiency loss due to the aluminum shield is less than 10\% for thicknesses below 100\,nm and X-ray energies above 1\,keV. We estimated an X-ray detection efficiency at 3.5\,keV of about 60\% to 65\%, which is mainly determined by the surface structures and not the aluminum shield. 

To increase the sensitivity in this range, a thinned back-illuminated CCD is needed. We simulated this scenario in Geant4, conservatively assuming that photons interacting in the partial charge collection layer are lost. We showed that the detection efficiency at 3.5\,keV can be increased to 85$\sim$90\%. For energies above 1.0\,keV, the efficiency loss introduced by the aluminum shield is $<$10\%. However, at lower energies, especially below the silicon L-shell at 99-150\,eV, the aluminum shield significantly reduces the detection efficiency. The efficiency loss due to data selection, pile-up, or reconstruction is not considered in this work. 

In this work, we demonstrated that thin aluminum coatings provide an effective shield against optical photons, suppressing skipper-CCD backgrounds in X-ray detection and in space-based instruments. Future work may include optimized fabrication processes, tests of back-illuminated thinned sensors, alternative shield materials, and the integration of shields into CCD arrays for astronomical and particle-physics applications. 


\section{Acknowledgments}

This manuscript was prepared using resources of the Fermi National Accelerator Laboratory (Fermilab) managed by Fermi Forward Discovery Group, LLC under Contract No. 89243024CSC000002 with the U.S.~Department of Energy, Office of Science, Office of High Energy Physics, and the DOE Early Career Award (DOE-ECA). Work performed at the Center for Nanoscale Materials, a U.S. Department of Energy Office of Science User Facility, was supported by the U.S. DOE, Office of Basic Energy Sciences, under Contract No. DE-AC02-06CH11357. The CCD development work was supported in part by the Director, Office of Science, of the U.S. Department of Energy under No. DE-AC02-05CH11231. The United States Government retains and the publisher, by accepting the article for publication, acknowledges that the United States Government retains a non-exclusive, paid-up, irrevocable, worldwide license to publish or reproduce the published form of this manuscript, or allow others to do so, for United States Government purposes.


\bibliography{references}{}
\bibliographystyle{aasjournal}

\end{document}